\documentclass[reprint,aps,prd,superscriptaddress,nofootinbib,floats,showpacs]{revtex4-2}
\usepackage[utf8]{inputenc}
\usepackage{graphicx}
\usepackage{amssymb}
\usepackage{textcomp}
\usepackage{amsmath}
\usepackage{tabularx}
\usepackage{bm}
\usepackage{times}
\usepackage{color}
\usepackage{ulem}
\usepackage{slashed}
\usepackage{multirow}
\usepackage{verbatim}
\usepackage{cancel}
\usepackage{subfigure}
\usepackage{epstopdf}
\usepackage{mathrsfs}
\usepackage{amsmath}
\usepackage{float}
\usepackage[export]{adjustbox}
\def\ba{\begin{align}}
\def\ea{\end{align}}
\usepackage{tikz}
\usetikzlibrary{shapes,snakes}
\numberwithin{equation}{section}
\usepackage[colorlinks=true, pdfstartview=FitV, linkcolor=blue, citecolor=blue, urlcolor=blue]{hyperref}
\allowdisplaybreaks[4]

\def\lsim{\mathrel{\raise.3ex\hbox{$<$\kern-.75em\lower1ex\hbox{$\sim$}}}}
\def\gsim{\mathrel{\raise.3ex\hbox{$>$\kern-.75em\lower1ex\hbox{$\sim$}}}}

\definecolor{orange}{rgb}{1,0.5,0}

\begin{document}

	\title{Effective dynamics of Janis-Newman-Winicour spacetime}
	
	\author{Faqiang Yuan}
	\affiliation{
		School of Physics and Astronomy, Key Laboratory of Multiscale Spin Physics (Ministry of Education), Beijing Normal University, Beijing 100875, China
	}

	\author{Shengzhi Li}
	\affiliation{
		School of Physics and Astronomy, Key Laboratory of Multiscale Spin Physics (Ministry of Education), Beijing Normal University, Beijing 100875, China
	}	
    \author{Zhen Li}
    \affiliation{
    School of Science, Jiangsu University of Science and Technology, Zhenjiang 212100, China
    }
	\author{Yongge Ma}
	\email{mayg@bnu.edu.cn}
	\affiliation{
		School of Physics and Astronomy, Key Laboratory of Multiscale Spin Physics (Ministry of Education), Beijing Normal University, Beijing 100875, China
	}
\begin{abstract}
The effective dynamics of the Janis-Newman-Winicour spacetime inspired by loop quantum gravity is studied. Two different schemes are considered to regularize the Hamiltonian constraint for the quantum dynamics. In the $\mu_0$ scheme in which the quantum parameters are treated as constants, the equations of motion generated by the effective Hamiltonian are solved analytically. The resulting quantum-corrected effective spacetime obviously extends the effective spacetime previously obtained in the literature. In the new effective spacetime, the naked singularity and the central singularity presented in the classical JNW spacetime are resolved by a series of quantum bounces. In the scheme of choosing the quantum parameters as Dirac observables, the effective dynamics is also solved in the light of the solution in $\mu_0$ scheme. It turns out that the resulting effective spacetime has singularities due to the appearance of the zero points of the time reparametrization functions. Hence, the effective theory in this scheme does not remain valid throughout the full spacetime.

\end{abstract}

\maketitle

\section{Introduction}
\label{sec:Intro}
Classical general relativity (GR) is known to predict the existence of spacetime singularities, which are generally expected to be resolved by certain quantum theory of gravity\cite{Penrose:1964wq,hawking2023large,wald2010general}. As one of the promising candidate theories of quantum gravity, Loop quantum gravity (LQG) is based on the canonical quantization of the holonomies of the connections and the fluxes of the densitized triads, which presents a picture of discrete spacetime at Planck scale\cite{ashtekar2004background,rovelli2004quantum,han2007fundamental,thiemann2008modern}. The geometric operators in LQG, such as the area operator, the volume operator and the length operator, generally have discrete spectra\cite{rovelli1995discreteness,ashtekar1997quantum,
	thiemann1998length,ma2010new}. It is reasonable to expect that the singularities in GR could be resolved by the quantum nature of spacetime geometry suggested by LQG. To test the ideas and the methods of LQG, its formal quantization prescriptions have been applied to symmetry-reduced models. The simplest theory that implements this concept is Loop Quantum Cosmology (LQC), which is constructed by applying the method of loop quantization to the homogeneous (and isotropic) cosmological models\cite{ashtekar2003mathematical,bojowald2008loop,
	ashtekar2009loop,ashtekar2011loop,banerjee2012introduction}. This theory provides a new perspective on the evolution of the early universe, where the big bang singularity in the classical theory is replaced by a quantum bounce\cite{ashtekar2006quantum0,ashtekar2006quantum2,
	ashtekar2008robustness,yang2009alternative}. The effective models of LQC play important roles in understanding the quantum theory by certain classical theory with quantum corrections. The idea to replace the classical Hamiltonian by a semiclassical effective Hamiltonian  works very well compared with the full quantum dynamics, even in the deep quantum regime\cite{ashtekar2011loop,li2023loop,
	agullo2023loop}. In LQC, there are two different quantization schemes, the so-called $\mu_0$ and $\bar{\mu}$ schemes, to regularize the Hamiltonian\cite{ashtekar2006quantum0,ashtekar2006quantum2}. Both of these two schemes can lead to the resolution of the classical singularity by a quantum bounce. The weakness of $\mu_0$ scheme is that the quantum bounce can occur even in the regions of low matter density\cite{ashtekar2006quantum0}. 

Besides the big-bang singularity, the other crucial issue in classical GR is the singularity inside black holes. The most extensively studied case for this issue is the interior of the Schwarzschild black hole. Since the spacetime in this case is isometric to that of the vacuum Kantowski-Sachs cosmological model, one can transport the techniques developed for homogeneous but anisotropic LQC to study this issue. It has been shown that the black hole singularity can be resolved by the effective dynamics of the model\cite{ashtekar2006quantum,modesto2006loop,boehmer2007loop,chiou2008phenomenological0,
corichi2016loop,olmedo2017black,ashtekar2018quantum,bodendorfer2019note,gan2024nonexistence,alesci2019quantum,alesci2020asymptotically,gan2022understanding,modesto2010semiclassical,yan2022constraints,liu2020shadow,zhu2020observational,ongole2022dirac}. There are two quantum parameters that should be fixed in this model. A few different schemes have been proposed for this issue. In the $\mu_0$ scheme, both the parameters are chosen as constants. However, the dynamical solution of the effective equation depends on the choice of the fiducial cell in this scheme\cite{ashtekar2006quantum,modesto2006loop}. In the $\bar{\mu}$ scheme, both the parameters are chosen as functions on the phase space. However, the quantum corrections to the classical spacetime may become large near the black hole horizon in this scheme, though this regime is expected to be classical\cite{boehmer2007loop,chiou2008phenomenological0}. In the scheme of the Ashtekar–Olmedo–Singh model, the quantum parameters are chosen as Dirac observables. Its effective dynamics not only avoid the dependence on the fiducial cell but also keep the classical regime near the horizon unchanged\cite{ashtekar2018quantum,bodendorfer2019note}.

To further study the issue of singularity resolution in LQG, we consider the homogeneous and spherically symmetric model of GR minimally coupled to a massless scalar field. As the interior of the Janis-Newman-Winicour (JNW) spacetime\cite{janis1968reality,
virbhadra1997janis,virbhadra2002gravitational,patil2012acceleration}, this model can also be regarded as the Kantowski-Sachs cosmology with a massless scalar field. Besides the central singularity, which is similar to that in the Schwarzschild case, there also exists another singularity known as the naked singularity in the classical model. The effective theory of this model was studied by two different schemes in Refs.\cite{chiou2008phenomenological,zhang2020quantum}. In the $\bar{\mu}$ scheme, the effective spacetime will descend into deep Planck regime after several bounces, where the semiclassical description can no longer be trusted\cite{chiou2008phenomenological}. In the scheme of choosing the quantum parameters as Dirac observables, both the central singularity and the naked singularity can be replaced by quantum bounces\cite{zhang2020quantum}. However, the resulting effective spacetime is incomplete and needs to be extended. In this paper, we will further study the effective theory of this model by two different schemes. In the $\mu_0$ scheme, we obtain the same equations of motion as those in Ref.\cite{zhang2020quantum}. However, the form of the solution to the effective dynamics will be given by choosing a positive lapse function, which is the extension needed in Ref.\cite{zhang2020quantum}. In the scheme of choosing the quantum parameters as Dirac observables, it will be shown that the singularities remain in the effective spacetime, in contrast to the result in Ref.\cite{zhang2020quantum}. 

The structure of this paper is as follows. In section II, the classical dynamics of the JNW spacetime will be reviewed. In section III, the effective dynamics of the JNW spacetime will be studied through two different schemes. Finally, the results will be summarized and discussed in section IV.

\section{Classical dynamics of Janis-Newman-Winicour spacetime}

\subsection{Hamiltonian theory}
    The action for GR minimally coupled to a massless scalar field reads
\begin{align}
	\label{action}
	S\left[g_{a b}, \phi\right]=\int \mathrm{d}^4 x \sqrt{-g}\left(\frac{1}{2\kappa} R-\frac{1}{2} g^{a b} \nabla_a \phi \nabla_b \phi\right) ,
\end{align}   
where the gravitational constant is denoted as $\kappa=8\pi G$, $g$ is the determinant of the spacetime metric $g_{ab}$, $R$ is the scalar curvature of $g_{ab}$, and $\phi$ is the scalar field. In the connection-dynamical formalism, the canonical variables for gravity consist of the $SU(2)$-connection $A_a^i$ and the densitized triad $E_j^b$\cite{ashtekar1987new}, and the those for the scalar field consist of the scalar $\phi$ and its momentum $\pi$. Their nontrivial Poisson brackets read 
\begin{align}
	\label{Poisson brackets}
	\left\{A_a^j(x), E_k^b(y)\right\}\!=\! \kappa \gamma \delta_a^b \delta_k^j \delta(x, y), \quad \left\{\phi(x), \pi_{\phi}(y)\right\}\!=\! \delta(x, y) ,
\end{align}
where $\gamma$ it the Barbero-Immirzi parameter\cite{barbero1994real,immirzi1997real,perez2006physical}. The total Hamiltonian of the system is composed of three first-class constraints, namely the Gauss constraint, the diffeomorphism constraint, and the Hamiltonian constraint\cite{zhang2020quantum}.

The spherically symmetric and static solution to such a system is the so-called  JNW metric given by\cite{janis1968reality,
	virbhadra1997janis,virbhadra2002gravitational,patil2012acceleration}
\begin{align}
	\label{JNW metric}
	\mathrm{d} s^2 &=-\left(1-\frac{B}{r}\right)^\nu d t^2+\left(1-\frac{B}{r}\right)^{-\nu} dr^2\nonumber \\
	&+r^2\left(1-\frac{B}{r}\right)^{1-\nu} d \Omega^2 
\end{align}
with the scalar field given by
\begin{align}
	\label{JNW scalar}
	\phi=\frac{q}{B \sqrt{4 \pi}} \ln(1-\frac{B}{r})
\end{align}
where $q$ denoted the scalar charge. The two parameters $B$ and $\nu$ are given by
\begin{align}
	\label{Bnu}
	\nu=\frac{2m}{B}, \quad B=2 \sqrt{m^2+q^2}
\end{align}
with $m$ standing for the Arnowitt-Deser-Misner (ADM) mass \cite{zhang2020quantum}. The exterior solution of (\ref{JNW metric}) satisfies $0<\nu<1$ and $r>B$. The scalar curvature $R=-\frac{B^2(\nu^2-1)}{2(r-B)^{2-\nu}r^{2+\nu}}$ diverge at $r=B$ where the singularity appears. By exchanging the coordinates of "time" and "space", the interior solution of Eq.~(\ref{JNW metric}) can be written as 
\begin{align}
	\label{JNW interior}
	\mathrm{d} s^2&=-\left(\frac{B}{\tau} -1\right)^{-\nu} d \tau^2+\left(\frac{B}{\tau} -1\right)^{\nu} d x^2 \nonumber \\
	&+\tau^2\left(\frac{B}{\tau} -1\right)^{1-\nu} d \Omega^2
\end{align}
with $0<\nu<1$ and $0<\tau<B$. Note that the central singularity locates at $\tau =0$ and the naked singularity locates at $\tau =B$. We are going to focus on the interior of JNW spacetime, which is homogeneous and spherically symmetric.

For the homogeneous and spherically symmetric model, the Cauchy slices $\Sigma$ have topology $\mathbb{R} \times \mathbb{S}^2$ and the symmetry group $\mathrm{S} \equiv \mathbb{R} \times SO(3)$. Since $\Sigma$ is non-compact in the $x$ direction, one needs to introduce an elementary cell $D \cong (0, L_0) \times \mathbb{S}^2$ in $\Sigma$ and restrict all integrals to this cell to avoid the divergence of integrations. By solving the Gaussian constraint, the gravitational connection and densitized triads take the form\cite{ashtekar2006quantum}
\begin{align}
	\label{symmetry variables}
	&A_a^i \tau_i \mathrm{~d} x^a  \!=\! \frac{c}{L_0} \tau_3 \mathrm{~d} x \!+\! b \tau_2 \mathrm{~d} \theta \!-\! b \tau_1 \sin \theta \mathrm{~d} \phi+\tau_3 \cos \theta \mathrm{~d} \varphi ,\\
	&E_i^a \tau^i \partial_a  =p_c \tau_3 \sin \theta \partial_x+\frac{p_b}{L_0} \tau_2 \sin \theta \partial_\theta-\frac{p_b}{L_0} \tau_1 \partial_\varphi,
\end{align}
and the scalar field and its conjugate momentum are reduced to the form
\begin{align}
	\label{reduced scalar}
	\phi(z)=\phi, \quad \pi_{\phi}(z)= \frac{\pi_{\phi}}{4\pi L_0} \sin\theta,
\end{align} 
where $\tau_i$ are $SU(2)$ generators related to the Pauli matrices $\sigma_i$ via $\tau_i = -\frac{i}{2}\sigma_i$, and $b, c, p_b, p_c, \phi, \pi_{\phi}$ represent the dynamical variables that are constants on a given $\Sigma$. The symplectic structure on phase space is reduced to
\begin{align}
	\label{reduced symplectic structure}
	\left\{b, p_b\right\}= \frac{\kappa \gamma}{8\pi} ,\quad \left\{c, p_c\right\}=\frac{\kappa \gamma}{4\pi}, \quad \left\{\phi, \pi_{\phi}\right\}=1 .
\end{align} 
The diffeomorphism constraint is satisfied automatically due to the homogeneity, and the Hamiltonian constraint is reduced to  
\begin{align}
	\label{reduced Hamiltonian}
	H \!=\! \int_D \mathrm{d}^3 x N C \!&=\!-\frac{4\pi}{\kappa \gamma^2} \frac{N \operatorname{sgn}\left(p_b\right) b}{\sqrt{\left|p_c\right|}}\! \left(\left(b \!+\! \frac{\gamma^2}{b}\right) p_b \!+\! 2 c p_c\right)\! \nonumber\\
	&+ \frac{N\pi_{\phi}^2}{8 \pi\left|p_b\right| \sqrt{p_c}} .
\end{align} 

\subsection{Classical dynamics}
    The classical equations of motion is determined by the Hamiltonian constraint (\ref{reduced Hamiltonian})
and symplectic structure (\ref{reduced symplectic structure}). After choosing the lapse function and writing down the corresponding Hamiltonian equations of motion, one obtains the solutions to the equations of motion\cite{zhang2020quantum}. 
Substituting this solutions to the general form of spherically symmetric metric,
\begin{align}
	\label{metric0}
	\mathrm{d} s^2=-N_t^2 \mathrm{~d} t^2+\frac{p_b^2}{\left|p_c\right| L_0^2} \mathrm{~d} x^2+\left|p_c\right|\left(\mathrm{d} \theta^2+\sin ^2 \theta \mathrm{~d} \varphi^2\right) ,
\end{align} 
one can obtain the explicit form of the spacetime metric. It should be noted that $c p_c =: \mathfrak{m}$ and $\pi_{\phi}$ are two Dirac observables which are constant along any classical dynamical trajectory, and they are related to $B$ and $\nu$ by $\pi_{\phi}^2=\frac{8 \pi^2 L_0^2 B^2 (1-\nu^2)}{\kappa}$ and $\mathfrak{m}=\frac{B \nu \gamma L_0}{2}$. 

It can be verify that the spacetime region around $t=t_0=\frac{1}{\kappa \gamma^2 L_0 B} \ln(\frac{1-\nu}{1+\nu})$ is classical by calculating the scalar curvature\cite{zhang2020quantum}. Hence, the dynamical fields at this moment can serve as an initial condition to match the solutions of the effective theory and the classical theory. A classical solution is given in Fig.1. It illustrates that both $p_b$ and $p_c$ monotonically decrease to zero toward both the past and future directions. This behavior results in the emergence of a naked singularity and a central singularity, respectively. With the solutions to the equations of motion, it is straightforward to check that the naked singularity occurs at\cite{zhang2020quantum}
\begin{align}
	\label{snaked singularity}
	\tau=B \Rightarrow t= - \infty \Rightarrow T= - \infty,
\end{align} 
and the central singularity occurs at
\begin{align}
	\label{center singularity}
	\tau=0 \Rightarrow t= \infty \Rightarrow T= - \infty .
\end{align}
\begin{figure}
	\centering
	\includegraphics[height=10cm,width=8cm]{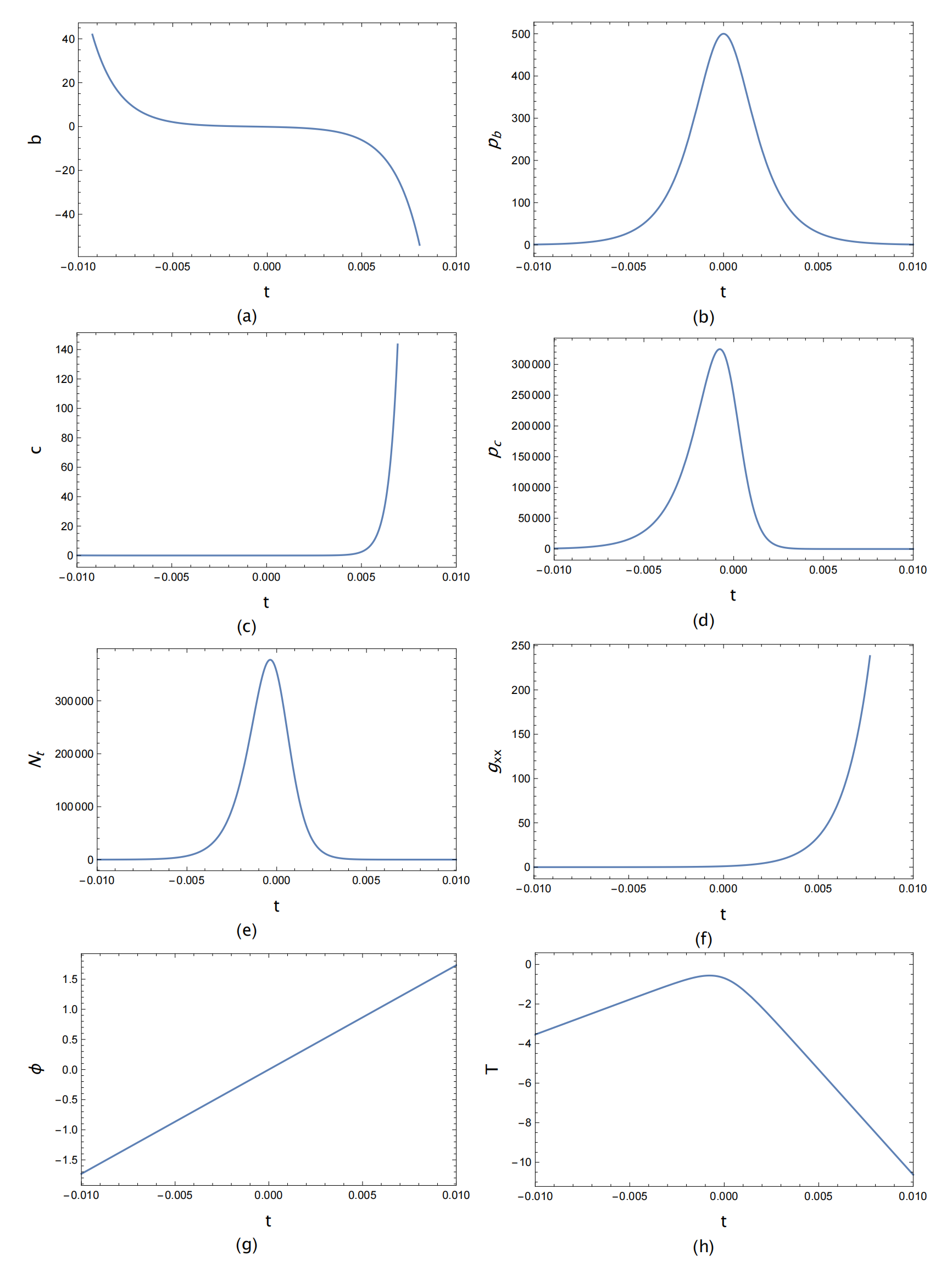}
	\caption{Plots of the classical evolutions for the variables (a) $b(t)$, (b) $p_b(t)$, (c) $c(t)$, (d) $p_c(t)$, and (e) $N_t(t)$, (f) $g_{xx}(t)$, (g) $\phi(t)$, and (h) $T(t)$: the parameters are chosen as $B = 10^3$, $\nu = 0.5$, $\kappa = 8\pi$, $\gamma=0.2375$, and $L_0=1$, and the initial data are chosen at $t_0=-0.000775$.}
\end{figure}

\section{Effective dynamics of Janis-Newman-Winicour spacetime}

The key idea in LQG is that the holonomy of a connection is the most fundamental object
at quantum level, rather than the connection itself. Hence, a naive way to get the effective theory of a LQC model is to replace the connections in the Hamiltonian by the holonomies and take the resulting Hamiltonian as the effective one\cite{ashtekar2011loop,li2023loop,
	agullo2023loop}. Thus, the effective Hamiltonian of the interior JNW spacetime can be obtained by replacing the connections by holonomies, i.e., by making the replacement $b\rightarrow \frac{\sin(\delta_b b)}{\delta_b}$ and $c\rightarrow \frac{\sin(\delta_c c)}{\delta_c}$ in the classical Hamiltonian. Here $\delta_b$ and $\delta_c$ are quantum parameters which can be chosen as constants or functions of the phase space variables. Different choices correspond to different quantization schemes and lead to different effective dynamics. Through this replacement, the effective Hamiltonian is obtained as 
\begin{align}
	\label{effective Hamiltonian1}
	H_{\mathrm{eff}}(N) \!&=\!-\!\frac{4\pi}{\kappa \gamma^2} \frac{N \! \operatorname{sgn} \! \left(p_b\right) \! \sin(\delta_b b) \!}{\sqrt{\left|p_c\right|} \delta_b}\! \Big( \!\Big( \! \frac{\sin(\delta_b b)}{\delta_b} \!+\! \frac{\gamma^2 \delta_b}{\sin(\delta_b b)} \!\Big)\! p_b \! \nonumber \\&+
	 2 \frac{\sin(\delta_c c)}{\delta_c} p_c \! \Big) \!+\! \frac{N\pi_{\phi}^2}{8 \pi\left|p_b\right| \! \sqrt{p_c} \!} .
\end{align} 
For later convenience, we choose a positive lapse $N_t=\kappa \gamma^2 |p_b| \sqrt{|p_c|}$ depending on the dynamical variables. With this choice of the lapse function, the effective Hamiltonian reads
\begin{align}
	\label{effective Hamiltonian2}
	H_{\mathrm{eff}}(N_t)&=  -4\pi \Big(\gamma^2 p_b^2+ \frac{\sin^2(\delta_b b)}{\delta_b^2} p_b^2\nonumber\\&+ 2 \frac{\sin(\delta_c c)}{\delta_c}p_c \frac{\sin(\delta_b b)}{\delta_b} p_b - \frac{\kappa \gamma^2 \pi_{\phi}^2}{32 \pi^2} \Big) .
\end{align}
We will consider the effective dynamics determined by this effective Hamiltonian and the symplectic structure (\ref{reduced symplectic structure}). In the quantization schemes introduced below, $\pi_\phi$ and $\frac{\sin(\delta_c c)}{\delta_c}p_c$ are constants of motion. For convenience later, we define the variables $y := \frac{\sin(\delta_b b)}{\delta_b} p_b$, $\mathfrak{g}:= \frac{\kappa \gamma^2 \pi_{\phi}^2}{32 \pi^2}$ and $\mathfrak{m}:=\frac{\sin(\delta_c c)}{\delta_c}p_c$. Here, $y$ and $\mathfrak{m}$ reduce to $b p_b$ and $c p_c$ in the classical theory in the limit $\delta_b \rightarrow 0, \delta_c \rightarrow0$. 

\subsection{Effective dynamics in the $\mu_0$ scheme}
 
In the $\mu_0$ scheme, the quantum regularization parameters of $\delta_b$ and $\delta_c$ are simply taken as constants and hence do not depend on the dynamical variables. It is straightforward to check that $\mathfrak{m}$ and $\mathfrak{g}$ are constants of motion in this scheme. 
In Ref.\cite{zhang2020quantum}, the lapse
$N_T=\gamma \operatorname{sgn}(p_b) \sqrt{|p_c|}\frac{\delta_b}{\sin(\delta_b b)}$ associated with the time coordinate $T$ was chosen to solve this dynamics. However, this lapse function diverges at $\sin(\delta_b b)=0$, leading to an incomplete effective quantum spacetime. In contrast, the lapse function chosen in this paper is positive and does not exhibit such singular behavior. Therefore, we expect to obtain an extension of the effective spacetime obtained in Ref.~\cite{zhang2020quantum}. The equation of motion of $y$ reads
\begin{align}
	\label{equation y}
	\frac{\mathrm{d}y}{\mathrm{d}t}=\left\{y, H_{\mathrm{eff}}(N_t)\right\}=-\kappa \gamma^3 p_b^2 \cos(\delta_b b) .
\end{align} 
By employing the effective Hamiltonian constraint $H_{\mathrm{eff}}(N_t) \approx0$, where the symbol $\approx$ means equaling on the constraint surface, we can solve $p_b^2$ as 
\begin{align}
	\label{p_b^2}
	p_b^2=\frac{1}{\gamma^2}(-y^2-2 \mathfrak{m}y +\mathfrak{g}) .
\end{align} 
Then Eq.~(\ref{equation y}) can be expressed as
\begin{align}
	\label{x}
	\frac{\mathrm{d}y}{\mathrm{d}t}
	=-\kappa\gamma^3  \mathrm{sgn}(\cos(\delta_b b))\sqrt{p_b^2}\sqrt{p_b^2-p_b^2 \sin^2(\delta_b b)} 
\end{align}
Substituting Eq.(\ref{p_b^2}) into Eq.(\ref{x}), we have
\begin{align}
	\label{x1}
	\frac{\mathrm{d}y}{\mathrm{d}t}
	=&-\kappa\gamma \mathrm{sgn}(\cos(\delta_b b)) \sqrt{-y^2-2 \mathfrak{m} y+ \mathfrak{g}} \nonumber \\&\quad \times \sqrt{-(1+\gamma^2 \delta_b^2)y^2-2 \mathfrak{m} y+ \mathfrak{g}}
\end{align} 
Taking the inverse of the above expression, one obtains
\begin{align}
	\label{equation t(y)}
	\frac{\mathrm{d}t}{\mathrm{d}y}=- \frac{1}{\kappa \gamma} \frac{\operatorname{sgn}(\cos(\delta_b b))}{\sqrt{-y^2-2 \mathfrak{m}y +\mathfrak{g}} \sqrt{-\bar{b} y^2 -2 \mathfrak{m} y +\mathfrak{g}}} ,
\end{align}  
where we denoted $\bar{b}:=1+\gamma^2 \delta_b^2$ for simplicity. Note that the quantities inside the square root in Eq.~(\ref{equation t(y)}) are always greater than or equal to zero since the first one come from $p_b^2$ and the second one come from $p_b^2 (1-\sin^2(\delta_b b))$. Note also that the sign function, $\operatorname{sgn}(\cos(\delta_b b))$, comes from the expression $\cos(\delta_b b)=\operatorname{sgn}(\cos(\delta_b b)) \sqrt{(1-\sin^2(\delta_b b))}$. For a given $\operatorname{sgn}(\cos(\delta_b b))$, we can immediately obtain the solution $t(y)$ by integrating the right-hand side of Eq.~(\ref{equation t(y)}) with respect to $y$. Since $\operatorname{sgn}(\cos(\delta_b b))$ may change between $1$ and $-1$ along the trajectory evolution, the solutions so obtained are local solutions valid for certain segments of the trajectory. To obtain a global solution to Eq.~(\ref{equation t(y)}), one has to know the value of $\operatorname{sgn}(\cos(\delta_b b))$ in each part of the trajectory. Hence, the next step is to analyze the range of $b$. 

The equation of motion of $b$ reads
\begin{align}
	\label{b equation}
	\frac{\mathrm{d}b}{\mathrm{d}t}\!=\! \left\{b, H_{\mathrm{eff}}(N_t)\right\} \!=\!-\kappa \gamma \!\left(\!\mathfrak{m} \frac{\sin \!(\!\delta_b b \!)\!}{\delta_b}\!+\! \frac{\sin^2 \!(\!\delta_b b \!)\!}{\delta_b^2} p_b \!+\! \gamma^2 p_b \!\right)\! .
\end{align}        
By employing the effective Hamiltonian constraint, we obtain
\begin{align}
	\label{mbequation}
	2\mathfrak{m} \frac{\sin(\delta_b b)}{\delta_b}=\frac{\mathfrak{g}}{p_b} -\frac{\sin^2(\delta_b b)}{\delta_b^2} p_b - \gamma^2 p_b .
\end{align}   
Substituting Eq.~(\ref{mbequation}) into Eq.~(\ref{b equation}), we have
\begin{align}
	\label{b equation2}
	\frac{\mathrm{d}b}{\mathrm{d}t}=-\frac{\kappa \gamma}{2}(\frac{\mathfrak{g}}{p_b} +\frac{\sin^2(\delta_b b)}{\delta_b^2} p_b + \gamma^2 p_b) .
\end{align}   
It should be noted that, as the triad component, $p_b(t)$ can not equal to zero. Let it be positive as in the classical theory. Then, the right hand side of Eq.~(\ref{b equation2}) can be estimated as 
\begin{align}
	\label{b equation3}
	-\frac{\kappa \gamma}{2}(\frac{\mathfrak{g}}{p_b} +\frac{\sin^2(\delta_b b)}{\delta_b^2} p_b + \gamma^2 p_b) &\leq -\frac{\kappa \gamma}{2}(\frac{\mathfrak{g}}{p_b} + \gamma^2 p_b) \nonumber \\
	&\leq - \kappa \gamma^2 \sqrt{\mathfrak{g}} .
\end{align}
Thus, the time derivative of $b$ is negative and has a upper bound since $\mathfrak{g}$ is a constant of motion. Therefore, we can conclude that $b \in (-\infty, \infty)$. Since $b$ is monotonic with respect to $t$, one can also take it as an internal dynamical time to describe the relational evolution.

Let us return to the equation of motion for $y$. Eq.~(\ref{equation t(y)}) can be divided into two sectors. In the sector of $\operatorname{sgn}(\cos(\delta_b b))=1$, i.e., $\delta_b b \in (-\frac{\pi}{2}+ 2 n \pi, \frac{\pi}{2}+ 2 n \pi)$ with $n \in \mathbb{Z}$, it becomes
\begin{align}
	\label{equation1 t(y)}
	\frac{\mathrm{d}t}{\mathrm{d}y}=- \frac{1}{\kappa \gamma} \frac{1}{\sqrt{-y^2-2 \mathfrak{m}y +\mathfrak{g}} \sqrt{-\bar{b} y^2 -2 \mathfrak{m} y +\mathfrak{g}}} .
\end{align}
In the sector of $\operatorname{sgn}(\cos(\delta_b b))=-1$, i.e., $\delta_b b \in (\frac{\pi}{2}+ 2 n \pi, \frac{3\pi}{2}+ 2 n \pi)$ it becomes
\begin{align}
	\label{equation2 t(y)}
	\frac{\mathrm{d}t}{\mathrm{d}y}=- \frac{1}{\kappa \gamma} \frac{-1}{\sqrt{-y^2-2 \mathfrak{m}y +\mathfrak{g}} \sqrt{-\bar{b} y^2 -2 \mathfrak{m} y +\mathfrak{g}}} .
\end{align}
The solutions of Eq.~(\ref{equation1 t(y)}) and Eq.~(\ref{equation2 t(y)}) are obtained respectively as
\begin{align}
	\label{solution1 t(y)}
	t(y)=\bar{t}(y)+ C_1[n] 
\end{align}
and
\begin{align}
	\label{solution2 t(y)}
	t(y)=-\bar{t}(y)+ C_2[n] ,
\end{align}
where $C_1[n]$ and $C_2[n]$ are integration constants and 
\begin{align}
	\label{tbar(y)}
	&\bar{t}(y)=-\frac{2}{\kappa \gamma} \sqrt{\frac{1}{L}} \nonumber\\
	&\times \mathrm{EllipticF}\left(\arcsin \big(\sqrt{\frac{M(\mathfrak{m}+\sqrt{\mathfrak{m}^2+\mathfrak{g}}+y)}{N(-\mathfrak{m}+\sqrt{\mathfrak{m}^2+\mathfrak{g}}-y)}} \big),\frac{P}{-L} \right) ,
\end{align}
with 
\begin{align}
	\label{tbar(y)c}
	&L:=2\sqrt{\mathfrak{m}^2+\mathfrak{g}} \sqrt{\mathfrak{m}^2+\bar{b}\mathfrak{g}}-2 \mathfrak{m}^2-\bar{b}\mathfrak{g}-\mathfrak{g}, \nonumber\\
	&M:=\mathfrak{m} -\mathfrak{m}\bar{b} +\bar{b}\sqrt{\mathfrak{m}^2+\mathfrak{g}} -\sqrt{\mathfrak{m}^2+\bar{b}\mathfrak{g}}, \nonumber\\
	&N:=-\mathfrak{m} +\mathfrak{m}\bar{b} +\bar{b}\sqrt{\mathfrak{m}^2+\mathfrak{g}} +\sqrt{\mathfrak{m}^2+\bar{b}\mathfrak{g}}, \nonumber\\ 
	&P:=2 \mathfrak{m}^2 +\mathfrak{g} +\bar{b}\mathfrak{g} +2\sqrt{\mathfrak{m}^2+\mathfrak{g}}\sqrt{\mathfrak{m}^2+\bar{b}\mathfrak{g}}
\end{align}
and $\mathrm{EllipticF}(u,k)$ being the first kind of elliptic integral (see its definition in Mathematica). One can also obtain the solution of $y$ with respect to $b$ as follows. Firstly, the solution of $p_b$ with respect to $b$ can be obtained by using the effective Hamiltonian constraint as
\begin{align}
	\label{pb(b)}
	p_b(b)=\frac{-\mathfrak{m} \frac{\sin(\delta_b b)}{\delta_b}+ \sqrt{(\mathfrak{m}\frac{\sin(\delta_b b)}{\delta_b})^2+(\gamma^2 +\frac{\sin^2(\delta_b b)}{\delta_b^2})\mathfrak{g}}}{\gamma^2 +\frac{\sin^2(\delta_b b)}{\delta_b^2}} .
\end{align}
Then, by definition we have
\begin{align}
	\label{y(b)}
	y(b)&=\frac{\sin(\delta_b b)}{\delta_b} \nonumber \\
	&\!\times \! \left(\! \frac{-\mathfrak{m} \frac{\sin(\delta_b b)}{\delta_b}+ \sqrt{(\mathfrak{m}\frac{\sin(\delta_b b)}{\delta_b})^2+(\gamma^2 +\frac{\sin^2(\delta_b b)}{\delta_b^2})\mathfrak{g}}}{\gamma^2 +\frac{\sin^2(\delta_b b)}{\delta_b^2}}\! \right)\! .
\end{align}
Next we can use the continuity of $t(b)$ to determine the integration constants $C_1[n]$ and $C_2[n]$. Suppose the initial data of $t=t_0$ be located at $b=0$. Firstly, in the region $\delta_b b \in (-\frac{\pi}{2},\frac{\pi}{2})$, Eq.~(\ref{solution1 t(y)}) gives $C_1[0]=t_0-\bar{t}(y(b=0))$. Then, for the region $\delta_b b \in (\frac{\pi}{2},\frac{3\pi}{2})$, the continuity of $t(b)$ at $\delta_b b=\pi/2$ implies $C_2[0]=2 \bar{t}(y(b=\frac{\pi}{2 \delta_b})) + C_1[0]$. Hence, all integration constants $C_1[n]$ and $C_2[n]$ can be determined by repeating the above procedure step by step. As a result, they are determined respectively as
\begin{align}
	\label{c1n}
	C_1[n]&=n\left[2\bar{t}(y(b=\frac{\pi}{2 \delta_b}))-2 \bar{t}(y(b=\frac{3\pi}{2 \delta_b}))\right]\nonumber\\
	&-\bar{t}(y(b=0)) +t_0 
\end{align}
and
\begin{align}
	\label{c2n}
	C_2[n]&=n\left[2\bar{t}(y(b=\frac{\pi}{2 \delta_b}))-2 \bar{t}(y(b=\frac{3\pi}{2 \delta_b}))\right] \nonumber \\
	&+2\bar{t}(y(b=\frac{\pi}{2 \delta_b}))-\bar{t}(y(b=0)) +t_0 .
\end{align}
Thus the solution $t(b)$ has been given by Eqs.~(\ref{solution1 t(y)}), (\ref{solution2 t(y)}) and (\ref{y(b)}).

    The equation of motion of $\phi$ reads
\begin{align}
	\label{phi equation}
	\frac{\mathrm{d}\phi}{\mathrm{d}t}=\left\{\phi, H_{\mathrm{eff}}(N_t)\right\}=\frac{\kappa \gamma^2 \pi_{\phi}}{4\pi} .
\end{align}
Since $\pi_{\phi}$ is a constant of motion, the solution for Eq.(\ref{phi equation}) can be obtained by fixing the initial value $\phi(t_0)=\sqrt{\frac{1-\nu^2}{2\kappa}} \ln(\frac{1-\nu}{1+\nu})$ as
\begin{align}
	\label{phi solution}
	\phi(t)=\frac{\kappa \gamma^2 \pi_{\phi}}{4\pi} t .
\end{align}
The solution can also be written in the form of $\phi(b)$ by using the function $t(b)$ obtained above.

    The equation of motion of $c$ reads
\begin{align}
	\label{cequation}
	\frac{\mathrm{d}c}{\mathrm{d}t}=\left\{c, H_{\mathrm{eff}}(N_t)\right\}=-2\kappa\gamma \frac{\sin(\delta_c c)}{\delta_c} y .
\end{align}
By Combining  Eqs. (\ref{equation t(y)}) and (\ref{cequation}), we obtain
\begin{align}
	\label{cequation2}
	\frac{\mathrm{d}\left[-\frac{1}{2} \ln(\tan \frac{\delta_c c}{2})\right]}{\mathrm{d}y}&\!=
	\frac{\mathrm{d}t}{\mathrm{d}y}\kappa\gamma y \nonumber\\
	&\!=\!-\! \frac{y\operatorname{sgn}(\cos(\delta_b b))}{\sqrt{\!-\!y^2\!-\!2 \mathfrak{m}y \!+\! \mathfrak{g}} \sqrt{\!-\! \bar{b} y^2 \!-\! 2 \mathfrak{m} y \!+\! \mathfrak{g}}} .
\end{align}
Then we have
\begin{align}
	&\int \mathrm{d}\left[-\frac{1}{2} \ln(\tan \frac{\delta_c c}{2})\right] \nonumber\\
	=&\int \mathrm{d}y \frac{-y\operatorname{sgn}(\cos(\delta_b b))}{\sqrt{\!-\!y^2\!-\!2 \mathfrak{m}y \!+\! \mathfrak{g}} \sqrt{\!-\! \bar{b} y^2 \!-\! 2 \mathfrak{m} y \!+\! \mathfrak{g}}}. \label{cequation3}
\end{align} 
The solution of Eq.(\ref{cequation3}) reads
\begin{align}
	\label{s0}
	&-\frac{1}{2} \ln(\tan \frac{\delta_c c}{2})+\lambda_0 \nonumber\\=&\int \mathrm{d}y \frac{-y\operatorname{sgn}(\cos(\delta_b b))}{\sqrt{\!-\!y^2\!-\!2 \mathfrak{m}y \!+\! \mathfrak{g}} \sqrt{\!-\! \bar{b} y^2 \!-\! 2 \mathfrak{m} y \!+\! \mathfrak{g}}}.
\end{align}
where $\lambda_0$ is the integration constant. For convenience, we define $T:=-\frac{1}{2} \ln(\tan \frac{\delta_c c}{2})+\lambda_0$. Taking its derivative, we obtain
\begin{align}
\label{s1}
\frac{\mathrm{d} c}{\mathrm{d} T}=-\frac{2\sin(\delta_c c)}{\delta_c}
\end{align}
This is precisely the equation of motion for $c$ when choosing the lapse $N_T$, i.e., the second equation in Eq.(3.3) in Ref.\cite{zhang2020quantum}. Therefore, the variable $T$ defined here is exactly the time coordinate corresponding to $N_T$. Solving the differential equation (\ref{s1}) requires fixing an integration constant $C^{(0)}$ through an initial condition. In Ref.\cite{zhang2020quantum}, this constant is determined by requiring that the value of $c$ at $T=0$ matches the classical result. For comparison, we fix the integration constant by $\lambda_0=\frac{1}{2} \ln(\frac{\gamma L_0 \delta_c}{B}\frac{\nu}{1-\nu^2}(\frac{1-\nu}{1+\nu})^{\nu})$ such that it corresponds to the value of $C^{(0)}$ chosen in Ref.\cite{zhang2020quantum}. Then the solution can be written as 
\begin{align}
	\label{csolution}
	\tan \left(\frac{\delta_c c(T)}{2}\right) =\frac{\gamma L_0 \delta_c}{B} \frac{\nu}{1-\nu^2}\left(\frac{1-\nu}{1+\nu}\right)^\nu e^{-2 T} .
\end{align}
Since $\mathfrak{m}$ is a constant of motion, the solution of $p_c$ can be obtained from Eq.(\ref{csolution}) as
\begin{align}
	\label{pcsolution}
	p_c(T) &=\frac{B^2}{4}\left(1-\nu^2\right)\left(\frac{1-\nu}{1+\nu}\right)^{-\nu}\nonumber\\
	&\times\left(e^{2 T}+\frac{\left(\gamma L_0 \delta_c\right)^2}{B^2} \frac{\nu^2}{\left(1-\nu^2\right)^2}\left(\frac{1-\nu}{1+\nu}\right)^{2 \nu} e^{-2 T}\right) .
\end{align} 
This give the solution in the form of Eq(3.8) in Ref.\cite{zhang2020quantum}. By using the time coordinate $T$, Eq.~(\ref{cequation2}) can be written as 
\begin{align}
	\label{Tequation}
	\frac{\mathrm{d}T}{\mathrm{d}y}=-\frac{y\operatorname{sgn}(\cos(\delta_b b))}{\sqrt{-y^2-2 \mathfrak{m}y +\mathfrak{g}} \sqrt{-\bar{b} y^2 -2 \mathfrak{m} y +\mathfrak{g}}},
\end{align}
which can be solved directly. The integration constant can be determined using the same method as solving $C_1[n]$ and $C_2[n]$. As a result, the solution is obtained as 
\begin{align}
	\label{T(b)1}
	T(y)=\bar{T}(y) \!+ \! D_1[n], \quad \mathrm{for} \quad \delta_b b \in (-\frac{\pi}{2}+ 2 n \pi, \frac{\pi}{2}+ 2 n \pi),  
\end{align}
\begin{align}
	\label{T(b)2}
	T(y)=\!-\! \bar{T}(y) \!+\! D_2[n], \quad \! \mathrm{for} \! \quad \delta_b b \in (\frac{\pi}{2}+ 2 n \pi, \frac{3\pi}{2}+ 2 n \pi), 
\end{align}
where the integration constants $D_1[n]$ and $D_2[n]$ are respectively given by
\begin{align}
	\label{D1n}
	D_1[n]&=n \left[2\bar{T}(y(b=\frac{\pi}{2\delta_b}))-2\bar{T}(y(b=\frac{3\pi}{2\delta_b}))\right]
	\nonumber\\
	&-\bar{T}(y(b=0))  
\end{align}
and
\begin{align}
	\label{D2n}
	D_2[n]&=n \left[2\bar{T}(y(b=\frac{\pi}{2\delta_b}))-2\bar{T}(y(b=\frac{3\pi}{2\delta_b}))\right]\nonumber\\
	&+2\bar{T}(y(b=\frac{\pi}{2\delta_b})) -\bar{T}(y(b=0))  .
\end{align}
Here the function $\bar{T}(y)$ reads
\begin{align}
	\label{Tbar(y)}
	&\bar{T}(y)=-2 \sqrt{\frac{1}{L}} \times (-\mathfrak{m}+\sqrt{\mathfrak{m}^2+\mathfrak{g}}) \nonumber\\
	&\times \mathrm{EllipticF}\Big(\arcsin \big(\sqrt{\frac{M(\mathfrak{m}+\sqrt{\mathfrak{m}^2+\mathfrak{g}}+y)}{N(-\mathfrak{m}+\sqrt{\mathfrak{m}^2+\mathfrak{g}}-y)}} \big),\frac{P}{-L} \Big)\nonumber\\
	&+4 \sqrt{\frac{1}{L}}  \times \sqrt{\mathfrak{m}^2+\mathfrak{g}} \nonumber\\
	&\! \times \! \mathrm{EllipticPi}\!\Big(\!\frac{N}{-M}, \arcsin \big(\sqrt{\frac{M(\mathfrak{m} \!+\! \sqrt{\mathfrak{m}^2\!+\! \mathfrak{g}}\!+\! y)}{N(\!-\! \mathfrak{m}\!+\! \sqrt{\mathfrak{m}^2 \!+\! \mathfrak{g}}\!-\! y)}}  \big), \frac{P}{\!-\! L} \! \Big)\! ,
\end{align}
where $\mathrm{EllipticPi}(l,u,k)$ is the third kind of elliptic integral (see its definition in Mathematica). Note that the combination of Eqs.~(\ref{T(b)1}), (\ref{T(b)2}) and (\ref{y(b)}) has implied the solution $T(b)$. Then the solutions $c(b)$ and $p_c(b)$ are obtained by substituting $T(b)$ into Eqs.~(\ref{csolution}) and (\ref{pcsolution}). It is easy to see that $p_c$ has a non-zero minimum value $p_c^{min}=\frac{1}{2}B \nu \gamma L_0 \delta_c$. Thus, we have obtained the solutions of the relative evolution for all dynamical  
variables with respect to the dynamical variable $b$.

To compare our solution with the solution obtained in Ref.\cite{zhang2020quantum}, we note that the lapse $N_T$ was chosen in Ref.\cite{zhang2020quantum} and the solutions for the dynamical variables was expressed as functions of $\xi:= \cos(\delta_b b)$. Thus the time coordinate ranges in a finite interval due to $\xi\in[-1, 1]$. Moreover, At the moment of $\xi=\pm1$, the lapse $N_T$ diverges and hence the scope of the solution is limited. In our treatment, the positive lapse $N_t=\kappa \gamma^2 |p_b|\sqrt{|p_c|}$ is chosen to solve the equations of motion. This lapse has no singularities. Thus, we can express the solutions of the dynamical variables as functions of $b$ with range of $b \in (-\infty, \infty)$. The solutions of $c(b)$, $p_c(b)$, $p_b(b)$, $g_{xx}(b)$, $t(b)$ and $T(b)$ are shown in Fig.2 respectively, where, as a comparison, the solutions obtained in Ref.\cite{zhang2020quantum} are also included. As shown in the figure, within the range of $\delta_b b \in [-\pi, \pi]$, our solution is the same as the solution of Ref.\cite{zhang2020quantum}, while our solution provides the necessary extension. In our solution, both $p_b$ and $p_c$ undergo a series of bounces toward both the past and future, so that the naked singularity and the central singularity are resolved respectively. As shown in Fig.2.(f), the time coordinate $T$ exhibits an oscillatory behavior with respect to $t$, which implies that $T$ cannot serve as a global time coordinate within this effective theory.

\begin{figure}
	\centering
	\includegraphics[height=7.5cm,width=8cm]{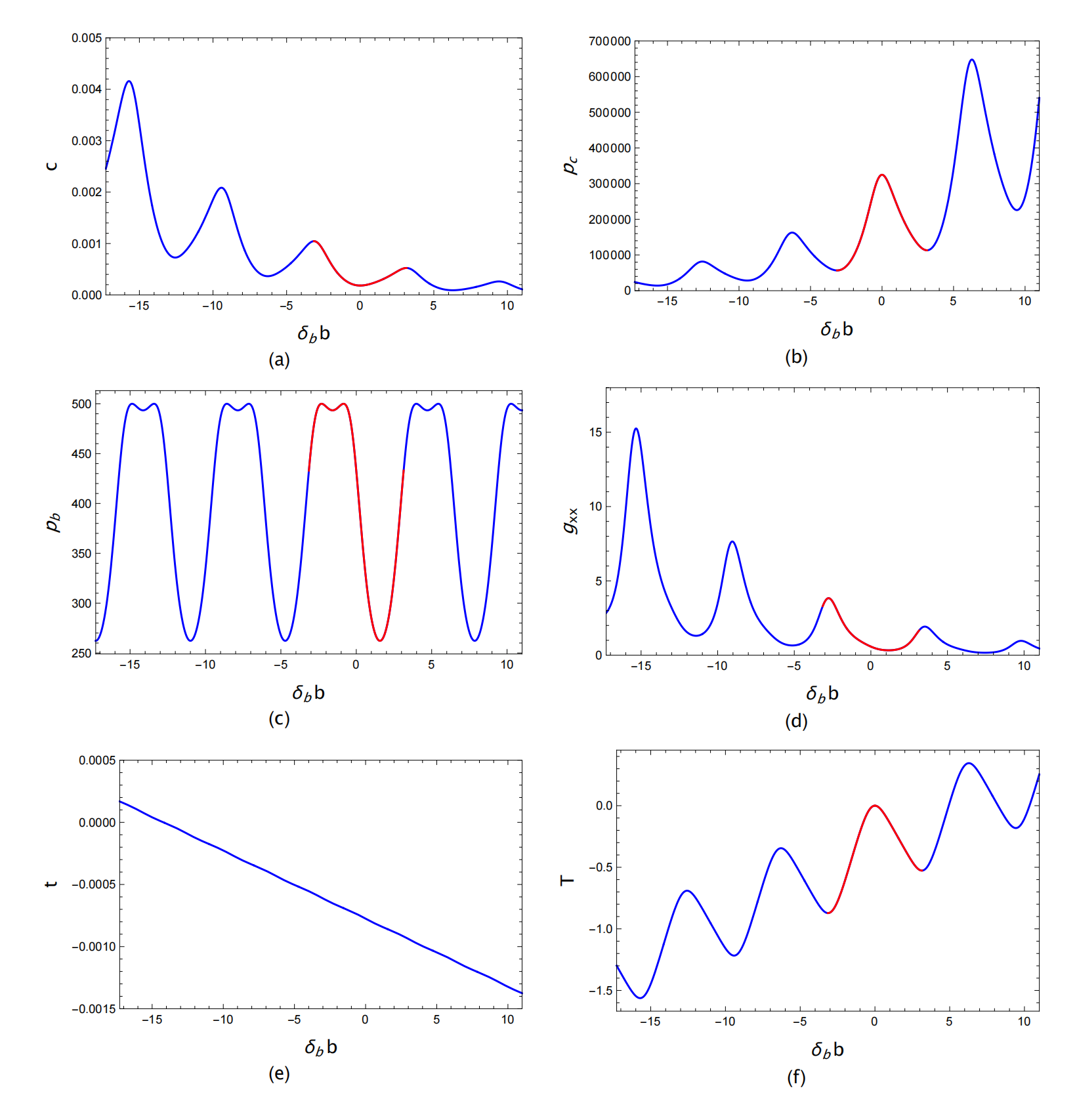}
	\caption{Plots of the evolutions in the effective theory with respect to $b$ for the variables  (a) $c(b)$, (b) $p_c(b)$, (c) $p_b(b)$, (d) $g_{xx}(b)$, (e) $t(b)$, and (f) $T(b)$: the parameters are chosen as $B =10^3$, $\nu = 0.5$, $\kappa = 8\pi$, $ \hbar=1$, and $L_0=1$. Here the blue lines represent our solutions, the red lines represent the solutions in Ref.\cite{zhang2020quantum}.}
\end{figure}

Let us consider the causal structure of the effective spacetime which we obtained. It is convenient to define the function 
\begin{align}
	\label{f(b)}
	f(b):= \frac{\mathrm{d}t}{\mathrm{d}b}=\frac{1}{-\kappa \gamma(\mathfrak{m} \frac{\sin(\delta_b b)}{\delta_b}+ \frac{\sin^2(\delta_b b)}{\delta_b^2} p_b + \gamma^2 p_b)} ,
\end{align}
where Eq.~(\ref{b equation}) is used. It is easy to see that $f(b)$ is a negative and upper bounded periodic function. Substituting our solution into Eq.~(\ref{metric0}), the effective spacetime metric can be written as
\begin{align}
	\label{metric1}
	\mathrm{d} s^2 =&-\kappa^2 \gamma^4 p_b^2(b) |p_c(b)| \mathrm{~d} t^2+\frac{p_b^2(b)}{\left|p_c(b)\right| L_o^2} \mathrm{~d} x^2 \nonumber\\ &+\left|p_c(b)\right|\left(\mathrm{d} \theta^2+\sin ^2 \theta \mathrm{~d} \varphi^2\right) \nonumber\\
	=&-\kappa^2 \gamma^4 p_b^2(b) |p_c(b)| f^2(b) \mathrm{~d} b^2+\frac{p_b^2(b)}{\left|p_c(b)\right| L_o^2} \mathrm{~d} x^2 \nonumber\\ &+\left|p_c(b)\right|\left(\mathrm{d} \theta^2+\sin ^2 \theta \mathrm{~d} \varphi^2\right) .
\end{align}
Note that both the integrals of $\int_0^{\infty} \mathrm{d} b \sqrt{-g_{00}(b)} $ and $\int_0^{-\infty} \mathrm{d} b \sqrt{-g_{00}(b)} $ are divergent. This indicates that, as the time parameter $b$ approaches positive or negative infinity from a finite value, the proper time diverges. Hence, the consequent spacetime can not be extended. It should also be noted that the effective spacetime metric is smooth and without singularity in the region of $b \in (-\infty, \infty)$. Therefore, this effective spacetime is geodesically complete and free of singularities.

To further understand the causal structure of the effective spacetime, we consider the expansions of the in-going and out-going radially null geodesics.
We employ the two null vectors tangent to the null geodesics as
\begin{align}
	\label{vectors}
	\ell^a_{ \pm}= \frac{1}{\sqrt{2}}(\sqrt{-g_{00}^{-1}}\partial^a b \pm \sqrt{g_{xx}^{-1}}\partial^a x) ,
\end{align}
satisfying $g_{ab}\ell^a_{+}\ell^b_{-}=-1$. The corresponding co-vectors read 
\begin{align}
	\label{covectors}
	\ell_a^{ \pm}= \frac{1}{\sqrt{2}}(-\sqrt{-g_{00}}\partial_a b \pm \sqrt{g_{xx}}\partial_a x) .
\end{align}
The expansions of these null geodesics are given by 
\begin{align}
	\label{expansion}
	\Theta^{ \pm}= S^{ab} \nabla_a \ell_b^{ \pm} &=\frac{\dot{g}_{\Omega \Omega}(b)}{\sqrt{2}\sqrt{-g_{00}(b)}g_{\Omega\Omega}(b)} \nonumber\\
	&=\frac{\dot{p_c}(b)}{\sqrt{2}\kappa \gamma^2 p_b(b) |f(b)| p_c^{\frac{3}{2}}(b)} ,
\end{align}
where the dot denotes the derivative with respect to $b$ and $S^{ab}$ is the projection map of $g^{ab}$ on the spatial 2-spheres perpendicular to $\ell^a_{ \pm}$. A 2-sphere is called a marginally trapped surface if $\Theta^{ \pm}=0$. Since the quantities, $p_b, f$ and $p_c$ remain finite for any finite $b$, Eq.~(\ref{expansion}) implies that a marginally trapped surface can exist if only if $\dot{p}_c(b) = 0$. A direct calculation gives
\begin{align}
	\label{pcdot}
	\dot{p}_c(b)=\frac{\mathrm{d} p_c}{\mathrm{d} t} \frac{\mathrm{d}t}{\mathrm{d} b}=\left\{p_c, H_{\mathrm{eff}}(N_t)\right\} f =2\kappa\gamma \cos(\delta_c c) p_c y f .
\end{align}
Since the quantities, $p_c$ and $f$ can not be zero, Eq.~(\ref{pcdot}) implies that a marginally trapped surface can exist at $\cos(\delta_c c)=0$ or $y=0$. The latter condition is equivalent to $b=n\frac{\pi}{ \delta_b}, n \in \mathbb{Z}$. Thus, there exist infinite numbers of marginally trapped surfaces in the effective spacetime. Across these marginally trapped surfaces, all the metric coefficients remain finite and non-zero, but $\dot{p_c}$ changes its sign. Thus, both $\Theta^{+}$ and $\Theta^{-}$ change signs simultaneously, and hence these marginally trapped surfaces are the transition surfaces from trapped region to anti-trapped region or from anti-trapped region to trapped region. Restricting to the first two dimensions, the line element (\ref{metric1}) can be written as
\begin{align}
	\label{metric2}
	\mathrm{d} s^2 =&-\kappa^2 \gamma^4 p_b^2(b) |p_c(b)| f^2(b) \mathrm{~d} b^2+\frac{p_b^2(b)}{\left|p_c(b)\right| L_o^2} \mathrm{~d} x^2 \nonumber\\
	=& \frac{p_b^2(b)}{\left|p_c(b)\right| L_o^2}(-\kappa^2 \gamma^4 L_0^2 p_c^2 f^2 \mathrm{~d} b^2+\mathrm{~d} x^2) \nonumber\\
	=& \frac{p_b^2(b)}{\left|p_c(b)\right| L_o^2}(-\mathrm{~d} \tilde{t}^2+\mathrm{~d} x^2) ,
\end{align}
where $\tilde{t}$ is defined by
\begin{align}
	\label{dtildet}
	\frac{\mathrm{d} \tilde{t}}{\mathrm{d} b}=\kappa \gamma^2 L_0 p_c(b) f(b),
\end{align}
and hence
\begin{align}
	\label{tildet}
	\tilde{t}(b)=\int_0^b \mathrm{d} h \kappa \gamma^2 L_0 p_c(h) f(h) .
\end{align}
Since $p_c f$ is negative and bounded above, one gets the range of $\tilde{t}$ as $\tilde{t} \in (-\infty,\infty)$. Employing the new coordinates defined by
\begin{align}
	\label{sk}
    \tilde{t}+x=\tan(\xi+\chi), \quad \tilde{t}-x=\tan(\xi-\chi) ,
\end{align}
the Penrose diagram of the metric (\ref{metric2}) is plotted as Fig.3.
\begin{figure}
	\centering
	\includegraphics[height=5.76cm,width=8cm]{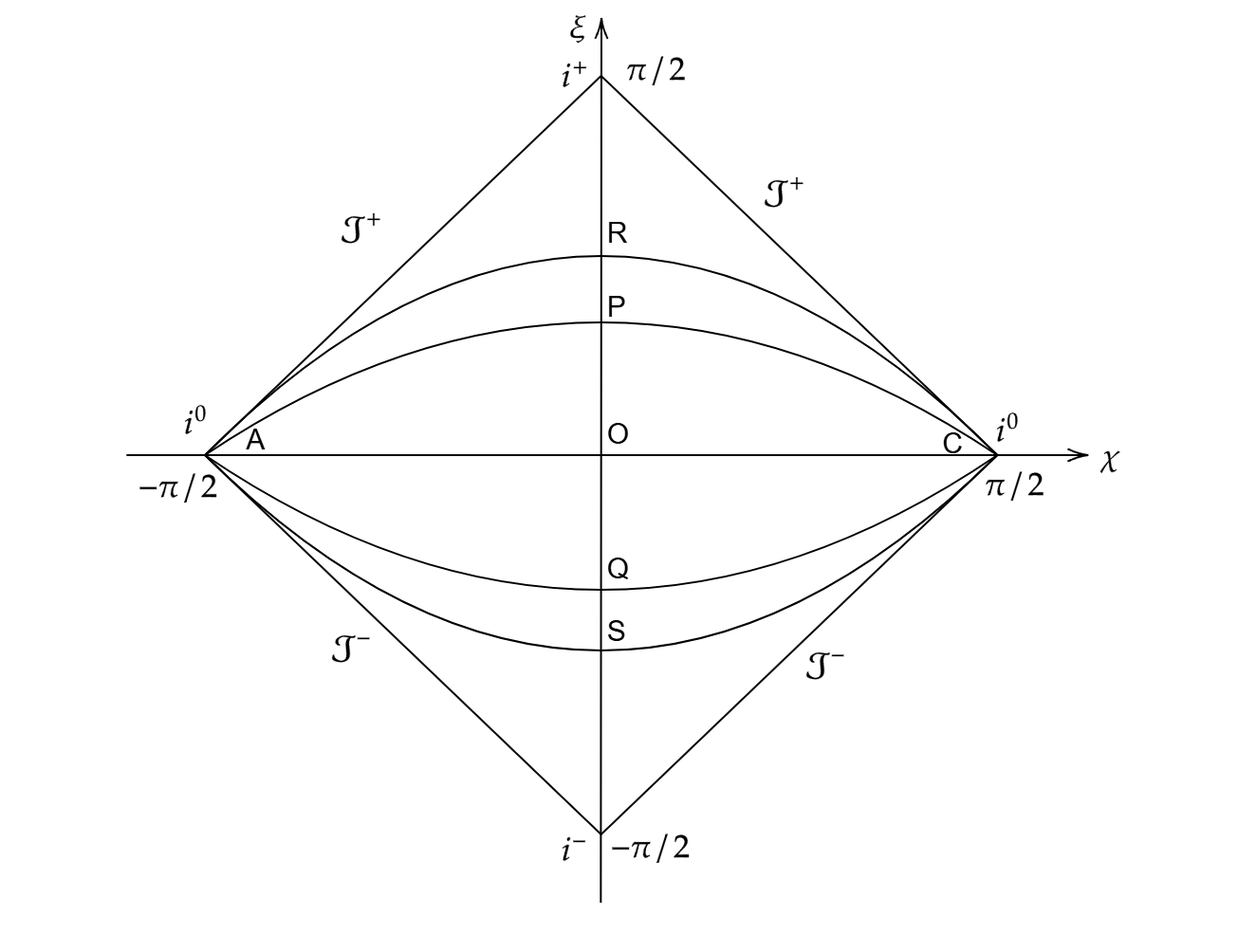}
	\caption{Plot of the Penrose diagram for the effective spacetime metric: the curve AOC corresponds to the transition surface at $b=0$. The curve APC, ARC, AQC, and ASC correspond to the transition surfaces at $b=-\frac{\pi}{\delta_b}$, $b=-\frac{2\pi}{\delta_b}$, $b=\frac{\pi}{\delta_b}$ and $b=\frac{2\pi}{\delta_b}$ respectively.}
\end{figure}

How to determine the values of the two quantum parameters $\delta_b $ and $\delta_c$ in $\mu_0$ scheme is a delicate issue. We propose to determine them by requiring that the resulting effective spacetime metric be independent of the choices of the fiducial cell's size $L_0$. According to Eq.(\ref{symmetry variables}), the connection component $b$ is invariant, while $c$ transforms as $c \rightarrow \alpha c$ under the rescaling $L_0 \rightarrow \alpha L_0$. In the classical theory, the resulting dynamics are independent of the rescaling of $L_0$. However, in the effective theory, $b$ and $c$ are replaced in the Hamiltonian constraint by $b\rightarrow \frac{\sin(\delta_b b)}{\delta_b}$ and $c\rightarrow \frac{\sin(\delta_c c)}{\delta_c}$, respectively. Hence, the trigonometric functions are introduced into the effective equations of motion. To ensure that the solutions of the effective equations remain independent of the choice of $L_0$, the term $ \frac{\sin(\delta_b b)}{\delta_b}$ and $ \frac{\sin(\delta_c c)}{\delta_c}$ should transform in the same way as $b$ and $c$ respectively under the rescaling of $L_0$. This requires that $\delta_b b$, $\delta_c c$, and $\delta_b$ remain invariant, while $\frac{1}{\delta_c}$ transforms in the same way as $c$. The latter condition implies that $\delta_c \propto L_0^{-1}$. In the homogeneous spherically symmetric model, one type of holonomies is the point holonomies obtained by integrating the connection along the longitudinal ($\theta$ direction) or equatorial ($\varphi$ direction) curves on the sphere\cite{ashtekar2006quantum}. In this case, $\delta_b$ represents the integration angle, and thus we can choose $\delta_b = 2\pi$. The another type of holonomies is the point holonomies obtained by integrating the connection along the $x$ directional curves. In this case, $\delta_c L_0$ represents the edge length of the integration\cite{ashtekar2006quantum}, and thus we can choose $\delta_c L_0 = \sqrt{\Delta}$, where $\Delta$ is the minimum nonzero eigenvalue of the area operator in LQG. It is straightforward to verify that this choice ensures that $\frac{\sin(\delta_b b)}{\delta_b}$ transforms in the same way as $b$, and $\frac{\sin(\delta_c c)}{\delta_c}$ transforms in the same way as $c$ under the rescaling of $L_0$. Thus, in contrast to the choice in Ref.\cite{ashtekar2006quantum}, our choice satisfies the requirement of ``fiducial cell independence''.

\subsection{Effective dynamics in the scheme of Dirac observables}
    To study the effective theory of black holes, the regularization 
scheme to choose the quantum parameters $\delta_b$ and $\delta_c$ as functions of the black 
hole mass, $m$, was proposed in Refs.\cite{ashtekar2018quantum,bodendorfer2019note}. In the case of Schwarzschild black holes, there are two Dirac observables $O_1(b,p_b,\delta_b)$ and $O_2(c,p_c,\delta_c)$, satisfying $O_1(b,p_b,\delta_b)=m=O_2(c,p_c,\delta_c)$ on the constraint surface. 
In general, the quantum parameter $\delta_b$ are taken to be arbitrary function of $O_1$, while $\delta_c$ are taken to be arbitrary function of $O_2$, i.e., $\delta_b=f_b(O_1(b,p_b,\delta_b))$ and $\delta_c=f_c(O_2(c,p_c,\delta_c))$\cite{bodendorfer2019note}. We will generalize this scheme to the effective theory of JNW spacetime. For convenience, we rewrite the effective Hamiltonian (\ref{effective Hamiltonian2}) as
\begin{align}
	\label{udeffective Hamiltonian}
	H_{\mathrm{eff}}(N_t) = 8\pi \gamma y L_0 (O_1-O_2) ,
\end{align}
where 
\begin{align}
	\label{O1}
    O_1:=\!-\! \frac{1}{2 \gamma}\left[\frac{\sin \left(\delta_b b\right)}{\delta_b} \!+\! \frac{\gamma^2 \delta_b}{\sin \left(\delta_b b\right)}\right] \frac{p_b}{L_o} \!+\! \frac{\kappa \gamma \delta_b \pi_{\phi}^2}{64\pi^2 L_0 \sin(\delta_b b) p_b} ,
\end{align}
\begin{align}
	\label{O2}
	O_2:=\left[\frac{\sin \left(\delta_c c\right)}{\gamma L_o \delta_c}\right] p_c .
\end{align}
Clearly, $O_1$, $O_2$ and $\pi_{\phi}$ are Dirac observables, and one has $O_1=O_2$ on the constraint surface. In the scheme of Dirac observables, we consider the quantum parameters as follows: $\delta_b$ as arbitrary function of $O_1$ and $\pi_{\phi}$ while $\delta_c$ as arbitrary function of $O_2$ and $\pi_{\phi}$, i.e.
\begin{align}
	\label{udscheme}
	\delta_b=f_b(O_1(b,p_b,\delta_b,\pi_\phi),\pi_\phi) ,
\end{align}
\begin{align}
	\label{udscheme2}
 \delta_c=f_c(O_2(c,p_c,\delta_c),\pi_\phi) .
\end{align}

Let us first focus on $\delta_b$. Since Eq.~(\ref{udscheme}) is an implicit function of $\delta_b$, $b$, $p_b$ and $\pi_{\phi}$, the sufficient condition for $\delta_b$ to be solvable as a function of $b$, $p_b$ and $\pi_{\phi}$, $\delta_b=\delta_b(b,p_b,\pi_{\phi})$, is
\begin{align}
	\label{exist condition}
	\frac{\partial \delta_b}{\partial \delta_b}- \frac{\partial f_b}{\partial \delta_b}=1-\frac{\partial f_b}{\partial O_1}\frac{\partial O_1}{\partial \delta_b} \neq 0 .
\end{align}
The equation of motion of $b$ reads
\begin{align}
	\label{udeb}
	\dot{b}=\left\{b, H_{\mathrm{eff}}(N_t)\right\}=\kappa \gamma^2 L_0 y \left(\frac{\partial O_1}{\partial p_b}+ \frac{\partial O_1}{\partial \delta_b} \frac{\partial \delta_b}{\partial p_b}\right) ,
\end{align}
where we used the Hamiltonian constraint in the second step. Note that $\delta_b$ depends on $p_b$ through an implicit function. Taking the partial derivative of both sides of the Eq.~(\ref{udscheme}) with respect to $p_b$, we get:
\begin{align}
	\label{udddbdpb}
	&\frac{\partial \delta_b}{\partial p_b}= \frac{\partial f_b}{\partial O_1}\left( \frac{\partial O_1}{\partial p_b} + \frac{\partial O_1}{\partial \delta_b} \frac{\partial \delta_b}{\partial p_b} \right) , \nonumber\\
	&\left(1- \frac{\partial f_b}{\partial O_1} \frac{\partial O_1}{\partial \delta_b} \right)\frac{\partial \delta_b}{\partial p_b}= \frac{\partial f_b}{\partial O_1} \frac{\partial O_1}{\partial p_b} , \nonumber\\
	&\frac{\partial \delta_b}{\partial p_b}= \frac{\frac{\partial f_b}{\partial O_1} \frac{\partial O_1}{\partial p_b}}{1- \frac{\partial f_b}{\partial O_1} \frac{\partial O_1}{\partial \delta_b}} .
\end{align}
Substituting the last equation of Eq.~(\ref{udddbdpb}) into Eq.~(\ref{udeb}), we obtain 
\begin{align}
	\label{udeb2}
	\dot{b}&=\kappa \gamma^2 L_0 y \frac{\partial O_1}{\partial p_b} \left(1+ \frac{\frac{\partial f_b}{\partial O_1} \frac{\partial O_1}{\partial \delta_b}}{1- \frac{\partial f_b}{\partial O_1} \frac{\partial O_1}{\partial \delta_b}}\right) \nonumber\\
	&=\kappa \gamma^2 L_0 y \frac{\partial O_1}{\partial p_b} \frac{1}{1- \frac{\partial f_b}{\partial O_1} \frac{\partial O_1}{\partial \delta_b}} .
\end{align}
It should be noted that Eq.~(\ref{udeb2}) is valid for $F_b\equiv1-\frac{\partial f_b}{\partial O_1}\frac{\partial O_1}{\partial \delta_b} \neq 0$. The equation of motion for $p_b$ reads
\begin{eqnarray}
	\label{udepb2}
	\begin{aligned}
		\dot{p_b}=-\kappa\gamma^2 L_0 y \left(\frac{\partial O_1}{\partial b} +\frac{\partial O_1}{\partial \delta_b} \frac{\partial \delta_b}{\partial b}\right).
	\end{aligned} 
\end{eqnarray}
Taking the partial derivative of both sides of the Eq. (\ref{udscheme}) with
respect to $b$, we get:
\begin{eqnarray}
	\label{udepb2x}
	\begin{aligned}
		\frac{\partial \delta_b}{\partial b}= \frac{\frac{\partial f_b}{\partial O_1} \frac{\partial O_1}{\partial b}}{1- \frac{\partial f_b}{\partial O_1} \frac{\partial O_1}{\partial \delta_b}}.
	\end{aligned} 
\end{eqnarray}
Substituting Eq.(\ref{udepb2x}) into Eq. (\ref{udepb2}), we obtain
\begin{align}
	\label{udeb2x}
	\dot{p_b}=-\kappa \gamma^2 L_0 y \frac{\partial O_1}{\partial b} \frac{1}{1- \frac{\partial f_b}{\partial O_1} \frac{\partial O_1}{\partial \delta_b}} .
\end{align}
Recall that in the $\mu_0$ scheme where $\delta_b$ is a constant, the equation of
motion for $b$ and $p_b$ under the same lapse function can be written respectively as
\begin{eqnarray}
	\label{e condition}
	\dot{b}=\kappa \gamma^2 L_0 y \frac{\partial O_1}{\partial p_b} ,
\end{eqnarray}
and
\begin{eqnarray}
	\label{pe condition}
	\dot{p_b}=-\kappa \gamma^2 L_0 y \frac{\partial O_1}{\partial b}.
\end{eqnarray}
Comparing Eqs.(\ref{udeb2}) and (\ref{udeb2x}) in the new scheme with Eqs.(\ref{e condition}) and (\ref{pe condition}) in the $\mu_0$ scheme, we find that the equations of motion for both $b$ and $p_b$ in the two schemes differ only by the additional factor $F_b$. Furthermore, Eq.(\ref{udeb2}) can be rewritten as
\begin{align}
	\frac{\mathrm{d} b}{\mathrm{d} t} F_b&=\kappa \gamma^2 L_0 y \frac{\partial O_1}{\partial p_b} \label{bre}, \\
	\frac{\mathrm{d} b}{\mathrm{d} t^\prime}=\frac{\mathrm{d} b}{\mathrm{d} t} \frac{\mathrm{d} t}{\mathrm{d} t^\prime}&=\kappa \gamma^2 L_0 y \frac{\partial O_1}{\partial p_b} \label{bre}.
\end{align}
where the new time parameter $t^\prime$ satisfies $\frac{\mathrm{d} t}{\mathrm{d} t^\prime}=F_b$. Hence,
for the $(b, p_b)$ sector, the equations of motion indicate that the
difference between the new scheme and the $\mu_0$ scheme can
be characterized by a certain time reparametrization. In the same way, one can obtain the equations of motion for $c$ and $p_c$ in the new scheme as
\begin{eqnarray}
	\label{uded2}
	\begin{aligned}
		\dot{c}=-2 \kappa \gamma^2 L_0 y \frac{\partial O_2}{\partial p_c} \frac{1}{1- \frac{\partial f_c}{\partial O_2} \frac{\partial O_2}{\partial \delta_c}},
	\end{aligned} 
\end{eqnarray}
and
\begin{eqnarray}
	\label{udepd2}
	\begin{aligned}
		\dot{p_c}=2 \kappa \gamma^2 L_0 y \frac{\partial O_2}{\partial c} \frac{1}{1- \frac{\partial f_c}{\partial O_2} \frac{\partial O_2}{\partial \delta_c}}.
	\end{aligned} 
\end{eqnarray}
Recall that in the $\mu_0$ scheme where $\delta_c$ is a constant, the equation of
motion for $c$ and $p_c$ under the same lapse function can be written respectively as
\begin{eqnarray}
	\label{u0ed2}
	\begin{aligned}
		\dot{c}=-2 \kappa \gamma^2 L_0 y \frac{\partial O_2}{\partial p_c} ,
	\end{aligned} 
\end{eqnarray}
and
\begin{eqnarray}
	\label{u0epd2}
	\begin{aligned}
		\dot{p_c}=2 \kappa \gamma^2 L_0 y \frac{\partial O_2}{\partial c} .
	\end{aligned} 
\end{eqnarray}
Comparing Eqs.(\ref{uded2}) and (\ref{udepd2}) in the new quantization scheme with Eqs.(\ref{u0ed2}) and (\ref{u0epd2}) in the $\mu_0$ scheme, we find that the equations of motion for both $c$ and $p_c$ differ only by the additional term $F_c:=1- \frac{\partial f_c}{\partial O_2} \frac{\partial O_2}{\partial \delta_c}\neq 0$. Hence, for the $(c, p_c)$
sector, the difference between the new scheme and the $\mu_0$ scheme can be characterized by another time reparametrization. Considering an arbitrary function $l$ on the phase space, its equation of motion gives
\begin{align}
	\label{emol}
	\frac{\mathrm{~d} l}{\mathrm{~d} t^\prime} \frac{\mathrm{~d} t^\prime}{\mathrm{~d} t}=\frac{\mathrm{~d} l}{\mathrm{~d} t}&=\{l,H(N_t)\}\nonumber\\
	&=\{l,\frac{N_t}{N_{t^\prime}} H(N_{t^\prime})\}\nonumber\\
	&=\frac{N_t}{N_{t^\prime}} \{l,H(N_{t^\prime})\}= \frac{\mathrm{~d} l}{\mathrm{~d} {t^\prime}} \frac{N_t}{N_{t^\prime}} ,
\end{align}
where in the second to the last step we have used the Hamiltonian constraint $H(N_{t^\prime})=0$. This shows that a time reparametrization of the solution corresponds to a change of the lapse function, and the reparametrized time and the lapse function satisfy
\begin{eqnarray}
	\label{emoTt}
	\frac{\mathrm{~d} {t^\prime}}{\mathrm{~d} t}= \frac{N_t}{N_{t^\prime}}.
\end{eqnarray}
Therefore, we can obtain the solutions of $(b, p_b, c, p_c)$ in the new scheme from the solutions in the $\mu_0$ scheme. The steps are as follows. Let $b_1$ be the time parameter corresponding to the lapse $N_1:=N_t F_b f$. Then Eqs. (\ref{b equation}), (\ref{f(b)}) and (\ref{udeb2}) implies that in the new scheme the evolution equations with respect to $b_1$ for the $(b,p_b)$ sector take the same form as their evolution equations with respect to $b$ in the $\mu_0$ scheme. Hence their solutions read $b(b_1)=b_1$ and $p_b(b_1)$ with the same functional form as in Eq.~(\ref{pb(b)}). Similarly, let $b_2$ be the time parameter corresponding to the lapse $N_2:=N_t F_c f$. Then the solutions for the $(c,p_c)$ sector in the new scheme, $c(b_2)$ and $p_c(b_2)$, have the same functional form as in Eqs.~(\ref{csolution}) and (\ref{pcsolution}). In order to obtain the solutions for the both $(b,p_b)$ and $(c,p_c)$ sectors in the new scheme with the same lapse $N_t$, we need to perform two time reparametrizations as follows:
\begin{align}
	\label{udt1t2}
	t_1(b_1)&=\int_{0}^{b_1} F_b(b) f(b) \mathrm{d}b +t_0, \nonumber\\
	t_2(b_2)&=\int_{0}^{b_2} F_c(b) f(b) \mathrm{d}b +t_0 ,
\end{align}
where $t_0$ is introduced to ensure the same initial time as the previous cases. By substituting the inverse functions of Eq.~(\ref{udt1t2}) back into the sectors of $(b,p_b)$ and $(c,p_c)$ respectively, we can obtain the solution $b(t_1^{-1}(t))$, $p_b(t_1^{-1}(t))$, $c(t_2^{-1}(t))$ and $p_c(t_2^{-1}(t))$ with the chosen lapse $N_t$, where $t_1^{-1}$ and $t_2^{-1}$ are the inverse functions of $t_1$ and $t_2$ respectively. 

The solution in this new scheme contains the inverse functions of $t_1$ and $t_2$. To determine the valid range of time parameter $t$ for the above solution, one needs to consider the monotonic intervals of the functions $t_1$ and $t_2$. Since $f(b)$ is a negative bounded function, it follows from Eq.(\ref{udt1t2}) that the properties of the functions $F_b(b)$ and $F_c(b)$ will determine the monotonic intervals. According to Eq.(\ref{udscheme}), in the expression of $F_b=1-\frac{\partial f_b}{\partial O_1}\frac{\partial O_1}{\partial \delta_b}$, the term $\frac{\partial f_b}{\partial O_1}$ is a function of $O_1$ and $\pi_{\phi}$ and thus a constant of motion. According to Eq.(\ref{udscheme2}), in the expression of $F_c=1- \frac{\partial f_c}{\partial O_2} \frac{\partial O_2}{\partial \delta_c}$, the term $\frac{\partial f_c}{\partial O_2}$ is a function of $O_2$ and $\pi_{\phi}$ and thus a constant of motion. According to Eq.(\ref{O1}), the term $\frac{\partial O_1}{\partial \delta_b}$ of $F_b$ can be calculated as
\begin{align}
	\label{uddo1ddb}
	\frac{\partial O_1}{\partial \delta_b}&=-\frac{1}{2\gamma} \frac{p_b}{L_0} \Big( \frac{b \cos(\delta_b b)}{\delta_b} -\frac{\sin(\delta_b b)}{\delta_b^2} +\frac{\gamma^2}{\sin(\delta_b b)} \nonumber\\
	&\quad-\frac{\gamma^2 \delta_b b \cos(\delta_b b)}{\sin^2(\delta_b b)}\Big) \nonumber\\ 
	&+\frac{\kappa \gamma \pi_{\phi}^2}{64\pi^2 L_0 p_b} \Big(\frac{1}{\sin(\delta_b b)} -\frac{\delta_b b \cos(\delta_b b)}{\sin^2(\delta_b b)}\Big) .
\end{align}
Substituting Eq.~(\ref{pb(b)}) into Eq.~(\ref{uddo1ddb}), one finds that $\frac{\partial O_1}{\partial \delta_b} \rightarrow -\infty$ as $b \rightarrow -\frac{\pi}{\delta_b}$, and  $\frac{\partial O_1}{\partial \delta_b} \rightarrow \infty$ as $b \rightarrow \frac{\pi}{\delta_b}$. From the expression of $F_b$, one can conclude that the range of $F_b$ is $(-\infty, \infty)$ in the case of $\frac{\partial f_b}{\partial O_1} \neq 0$. Therefore, there must exist a $b_0$ such that $F_b(b_0)=0$. 
The relevant term of $F_c$ can be calculated as
\begin{align}
	\label{uddo2ddc}
	\frac{\partial O_2}{\partial \delta_c}= \frac{c \cos(\delta_c c) p_c}{\gamma L_0 \delta_c} - \frac{\sin(\delta_c c) p_c}{\gamma L_0 \delta_c^2} .
\end{align}
Substituting Eqs.~(\ref{csolution}), (\ref{pcsolution}), (\ref{T(b)1}), (\ref{T(b)2}) and (\ref{y(b)}) into Eq.~(\ref{uddo2ddc}), one finds that $\frac{\partial O_2}{\partial \delta_c} \rightarrow - \infty$ as $b \rightarrow -\infty$, and $\frac{\partial O_2}{\partial \delta_c} \rightarrow 0$ as $b \rightarrow \infty$. From the expression of $F_c$, one can conclude that the range of $F_c$ contains $(-\infty, 1)$ in the case of $\frac{\partial f_b}{\partial O_1} < 0$. This implies that there must exist a $b^{\prime}_0$ such that $F_c(b^{\prime}_0)=0$ in this case. However, in the case of $\frac{\partial f_c}{\partial O_2} > 0$, the result is indeterminate and a case-by-case analysis is necessary. The functions $F_b$, $F_c$, $t_1$ and $t_2$ are plotted in Fig.4. It is shown that the function $F_b$ always has a zero point both for $\frac{\partial f_b}{\partial O_1} > 0$ and $\frac{\partial f_b}{\partial O_1} < 0$, whereas the function $F_c$ has a zero point for $\frac{\partial f_c}{\partial O_2} > 0$ but no zero point for $\frac{\partial f_c}{\partial O_2} < 0$. The appearance of these zero points implies that the time reparametrization function $t_1$ or $t_2$ in Eq.~(\ref{udt1t2}) are not monotonic.
\begin{figure}
	\centering
	\includegraphics[height=10cm,width=8cm]{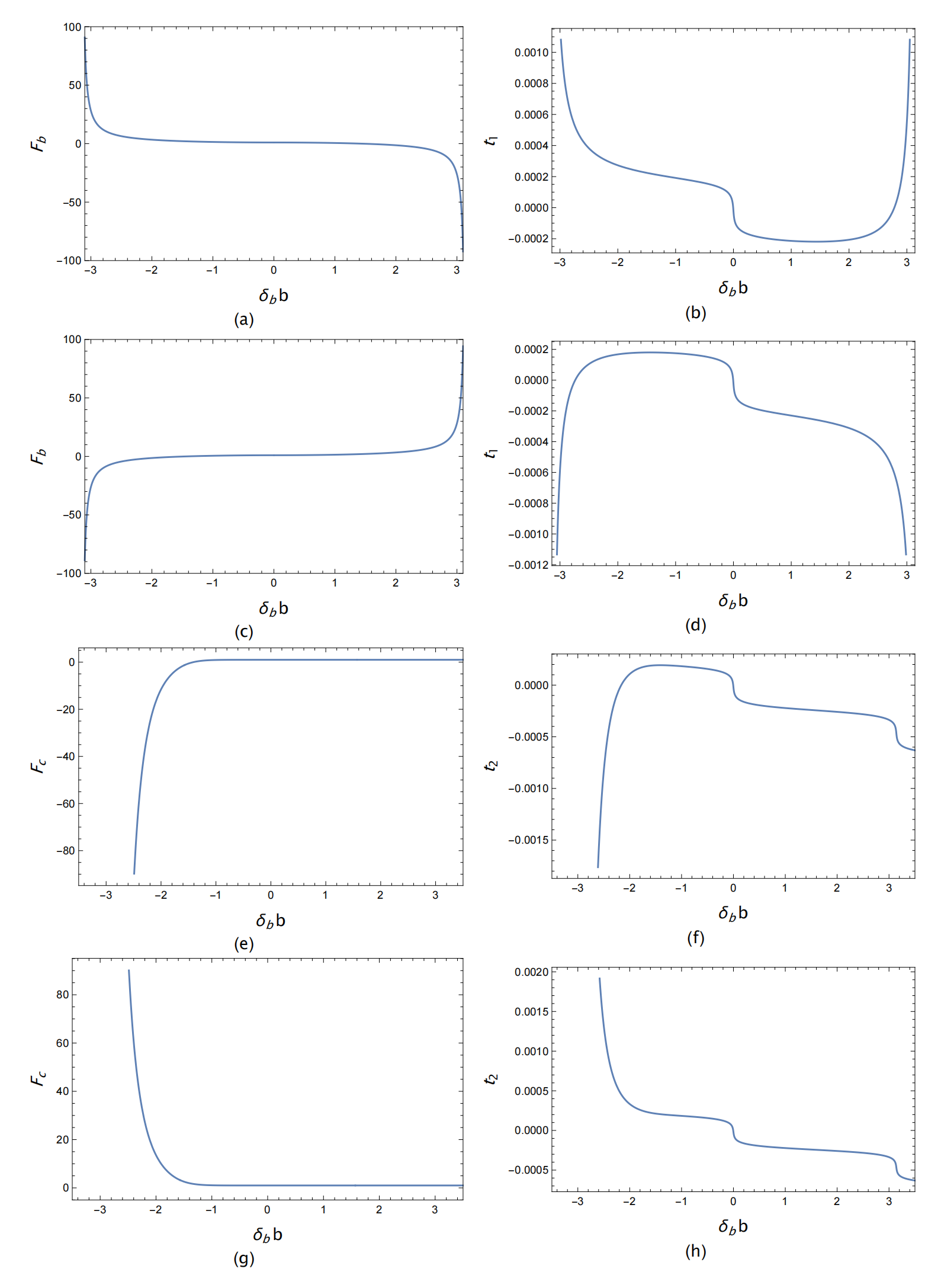}
	\caption{Plot of $F_b$, $F_c$, $t_1$ and $t_2$ with the parameters $B =  4\times 10^4$, $\nu = 0.5$, $\kappa = 8\pi$, $\hbar=1$, $\gamma=0.2375$, $L_0=1$, $\delta_b=0.034$, $\delta_c=0.0374$: the constants of motion are fixed as $\frac{\partial f_b}{\partial O_1}=2.108 \times 10^{-6} >0$ in (a) and (b) while $\frac{\partial f_b}{\partial O_1}=-2.108 \times 10^{-6} <0$ in (c) and (d), $\frac{\partial f_c}{\partial O_2}=-8.386 \times 10^{-6} <0$ in (e) and (f) while $\frac{\partial f_c}{\partial O_2}=8.386 \times 10^{-6} >0$ in (g) and (h).}
\end{figure}

As shown below, the occurrence of zero points in $F_b$ or $F_c$ will lead to singularities. The curvature scalar of the spacetime metric can be calculated as
\begin{align}
	\label{udcR}
	R&=\frac{4 \gamma^4 \kappa^2 p_b^4 p_c^2- 4p_c^2 \dot{p}_b^2- 4 p_b p_c \dot{p}_b \dot{p}_c- p_b^2 \dot{p}_c^2}{2 \gamma^4 \kappa^2 p_b^4 p_c^3} \nonumber\\
	&+ \frac{4 p_b p_c^2 \ddot{p}_b +2p_b^2  p_c \ddot{p}_c}{2 \gamma^4 \kappa^2 p_b^4 p_c^3}.
\end{align}
In the new scheme, the solutions for $p_b$ and $p_c$ are reparametrizations of those obtained in the $\mu_0$ scheme. Since $p_b$ and $p_c$ remain finite throughout the evolution in the $\mu_0$ scheme, they are consequently finite in the new scheme as well. The equations of motion of $p_b$ and $p_c$ with respect to $N_t$ read respectively
\begin{align}
	\label{uddpb}
	\dot{p_b}=\frac{1}{F_b} \left(\kappa \gamma \cos(\delta_b b) p_b(y+\mathfrak{m})\right) ,
\end{align}
and
\begin{align}
	\label{uddpc}
	\dot{p_c}=\frac{1}{F_c} \left(2\kappa \gamma \cos(\delta_c c) p_c y\right) .
\end{align}
Note that $F_b=0$ and $F_c=0$ respectively lead to divergences in $\dot{p_b}$ and $\dot{p_c}$
. Since both $p_b$ and $p_c$ in the denominators of Eq.~(\ref{udcR}) always take finite values, the curvature scalar $R$ diverges at these zero points of $F_b=0$ and $F_c=0$. Therefore, the occurrence of zero points in $F_b$ or $F_c$ indicates the presence of singularities in the spacetime. 

As a result, the signs of the two constants of motion, $\frac{\partial f_b}{\partial O_1}$ and $\frac{\partial f_c}{\partial O_2}$, determine whether the effective spacetime exhibits singularities in the new scheme. Specifically, the case of $\frac{\partial f_b}{\partial O_1} \neq 0$ or $\frac{\partial f_c}{\partial O_2} < 0$ will lead to a singular spacetime in the new scheme. The case of $\frac{\partial f_b}{\partial O_1} = 0$ and $\frac{\partial f_c}{\partial O_2} > 0$ requires further investigation for specific $f_c$. 
For the case of $\frac{\partial f_b}{\partial O_1} = 0$ and $\frac{\partial f_c}{\partial O_2} = 0$, such as $\delta_b=2\pi$ and $\delta_c= \frac{1}{\pi_{\phi}}$, the solution of spacetime metric has the same form as in $\mu_0$ scheme and the effective spacetime has no singularity. 

We now analyze a specific example by considering the following expressions for $\delta_b$ and $\delta_c$ given in Eqs.(4.13) and (4.14) of Ref.\cite{zhang2020quantum},
\begin{align}
	\label{1db}
	\delta_b =\left(\frac{\sqrt{\Delta}}{\sqrt{2 \pi}\left(\frac{\nu+1}{2} \gamma\right)^{\nu+1} \frac{B}{2^\nu}}\right)^{\frac{1}{\nu+2}} \equiv g_b(B,\nu)
\end{align}
and
\begin{align}
	\label{1dc}
	\delta_c  = \frac{1}{L_0 \nu}\left(\left(\frac{(1+\nu)^2 \Delta}{8 \pi}\right)^{\frac{1+\nu}{\nu}} \frac{\gamma}{4 B}\right)^{\frac{\nu}{\nu+2}} \equiv g_c(B,\nu),
\end{align}
where the Dirac observables $B$ and $\nu$ are given by $B=2\sqrt{m^2+\frac{\kappa \pi_{\phi}^2}{32\pi^2 L_0^2}}$ and $\nu=m /\sqrt{m^2+\frac{\kappa \pi_{\phi}^2}{32\pi^2 L_0^2}}$. Thus, in the new scheme, the specific quantum parameters are chosen as 
\begin{align}
	\label{2db}
	\delta_b &=f_b(O_1,\pi_{\phi})\nonumber\\
	&=g_b \left(2\sqrt{O_1^2+\frac{\kappa \pi_{\phi}^2}{32\pi^2 L_0^2}},O_1 /\sqrt{O_1^2+\frac{\kappa \pi_{\phi}^2}{32\pi^2 L_0^2}}\right)
\end{align}
and
\begin{align}
	\label{2dc}
	\delta_c  &= f_c(O_2,\pi_{\phi})\nonumber\\
	&=g_c \left(2\sqrt{O_2^2+\frac{\kappa \pi_{\phi}^2}{32\pi^2 L_0^2}},O_2 /\sqrt{O_2^2+\frac{\kappa \pi_{\phi}^2}{32\pi^2 L_0^2}}\right).
\end{align}
A straightforward computation yields that the two constants of motion satisfy $\frac{\partial f_b}{\partial O_1} > 0$ and $\frac{\partial f_c}{\partial O_2} < 0$ in this case. Therefore, the effective spacetime must be a singular spacetime. The subplots (a), (b), (e) and (f) of Fig.4 are the plots of the functions $F_b$, $t_1$, $F_c$ and $t_2$ in this scheme respectively. The corresponding evolutions of $b$, $p_b$, $c$, $p_c$, $\dot{p}_c$, $\dot{p}_b$ and $R$ are plotted in Fig.5. These plots indicate $F_b \approx 0$ at $t \approx -0.0002189$, which leads to a sharp increase in $|\dot{p}_b|$. Since both $p_b$ and $p_c$ remain finite at this moment, this results in a sharp increase in $|R|$. Similarly, because of $F_c \approx 0$ at $t \approx 0.0001926$, a sharp increase in $|\dot{p}_c|$ appears and in turn causes a sharp increase in $|R|$. Therefore, the resulting spacetime possesses two singularities. 
\begin{figure}
	\centering
	\includegraphics[height=10cm,width=8cm]{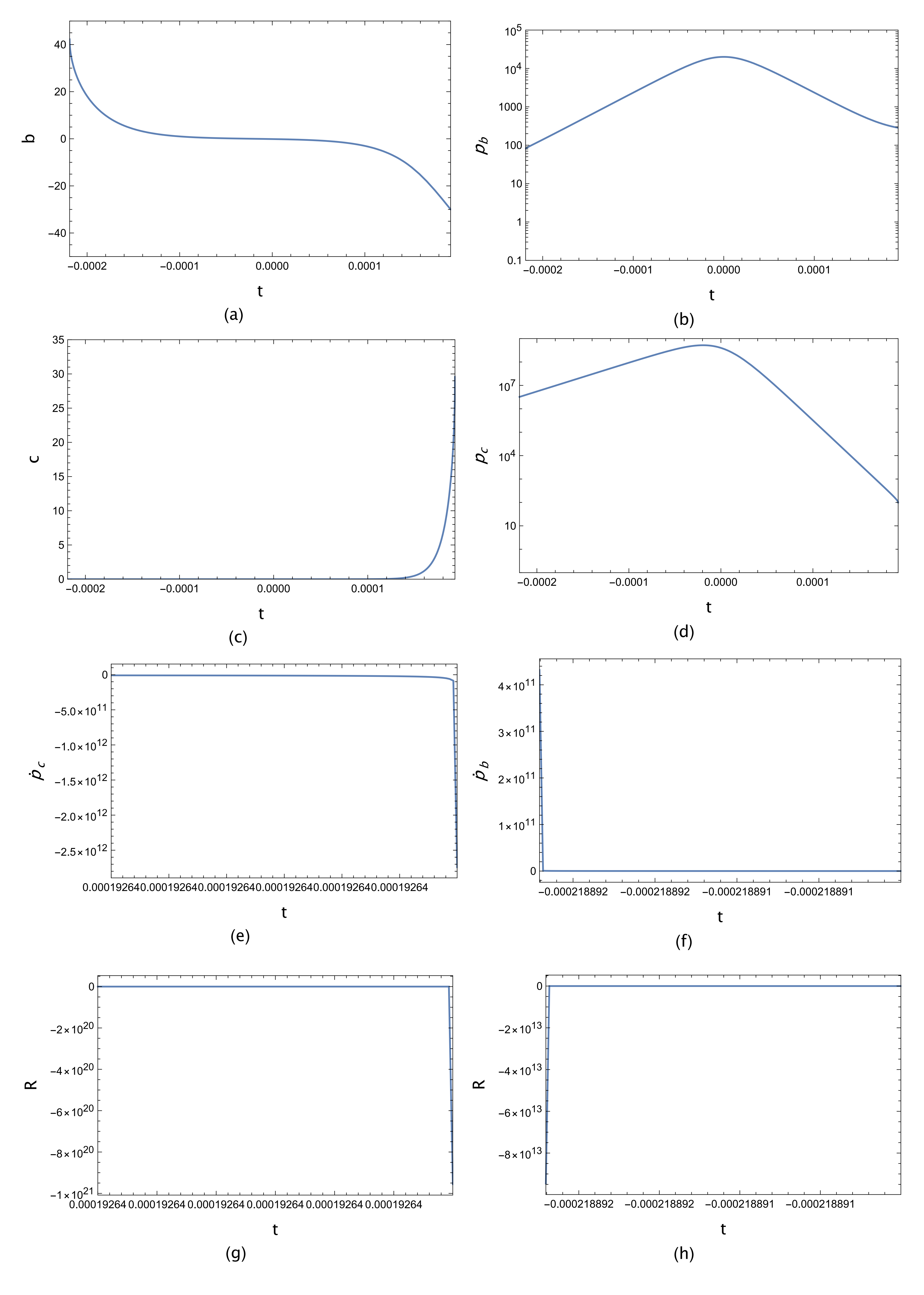}
	\caption{Plots of (a) $b(t)$, (b) $p_b(t)$, (c) $c(t)$, (d) $p_c(t)$, (e) $\dot{p}_c(t)$, (f) $\dot{p}_b(t)$, (g) and (h) $R(t)$ with the parameters $B = 4\times 10^4$, $\nu = 0.5$, $\kappa = 8\pi$, $\hbar=1$, $\gamma=0.2375$, $L_0=1$, $\delta_b=0.034$, $\delta_c=0.0374$: the constants of motion are fixed as $\frac{\partial f_b}{\partial O_1}=2.108 \times 10^{-6}$ and $\frac{\partial f_c}{\partial O_2}=-8.386 \times 10^{-6}$ such that both $F_b$ and $F_c$ have zero points, and these zero points lead to the two sharp increases of $|R|$ shown in (g) and (h) respectively.}
\end{figure}

\section{Summary and Discussion}
In previous sections, the effective dynamics of JNW spacetime inspired by LQG has been studied. In the $\mu_0$ scheme, we solved the equations of motion for the dynamical variables analytically. The resulting solution has been expressed as functions of the dynamical variable $b$ which was proved to be suitable as a time parameter. Thus, the effective spacetime proposed in Ref.\cite{zhang2020quantum} has been obviously extended. The solutions for $p_b$ and $p_c$ indicate that the effective spacetime undergoes a series of bounces which resolve the naked singularity and the central singularity presented in the classical JNW spacetime. Based on the solutions of dynamical variables, we constructed the quantum-corrected effective metric. Our analysis shows that there are infinite numbers of transition surfaces from trapped region to anti-trapped region or anti-trapped region to trapped region in this effective spacetime. The spacetime has no singularity and the effective dynamics is valid for the entire spacetime. 

The second scheme is to choose the quantum parameters as Dirac observables. By generalizing the method proposed in Ref.\cite{bodendorfer2019note} for the Schwarzschild black holes, we obtained the solutions of the effective dynamics for the $(b, p_b)$ and $(c, p_c)$ sectors separately from the solution in the $\mu_0$ scheme. As a result, the whole solution in this new scheme was obtained by performing time reparametrizations of the separate solutions $b$, $p_b(b)$ and $c(b)$, $p_c(b)$. It turns out that the resulting effective spacetime has singularities due to the appearance of the zero points of the two functions $F_b$ and $F_c$ in the expression of the solution. It was also shown by an example that the JNW effective spacetime does exhibit singularities. Hence this effective theory does not remain valid throughout the full spacetime.

While the effective dynamics of JNW spacetime in the two schemes has been deeply studied in this paper, there are still a few related issues deserving further investigation. For example, in the scheme of choosing the quantum parameter as Dirac observables, one may consider a more general setting of $\delta_b=f_1(O_1,O_2,\pi_{\phi})$ and $\delta_c=f_2(O_1,O_2,\pi_{\phi})$. One may also consider a hybrid scheme in which the quantum parameters $\delta_b$ and $\delta_c$ are treated in different schemes as was done in certain vacuum spherically symmetric models\cite{gan2025new}. Moreover, it is desirable to extend the methods developed in this paper to the model of GR non-minimally coupled to a scalar field. These open issues are left for our further investigations. 

\section{Acknowledgments}
\label{a}
This work is supported by the National Natural Science Foundation of China (Grant No.12275022). We thank Cong Zhang, Weilin Mu for helpful discussions.

\bibliographystyle{amsalpha}

\end{document}